\documentclass[conference]{IEEEtran}
\usepackage[utf8]{inputenc}
\usepackage[T1]{fontenc}
\usepackage[pdftex]{graphicx}
\usepackage{graphicx}
\usepackage{subfig}
\usepackage{makecell}
\usepackage{tikz}
\usepackage{pgfplots} 
\usetikzlibrary{decorations.text}
\usepackage{ifthen}
\usetikzlibrary{spy,calc}
\usepackage{courier}
\usepackage{arydshln}
\usepackage{amsmath}
\usepackage{cases}
\usepackage{multirow}
\usepackage{float}
\usepackage{verbatim}
\usepackage{booktabs} 
\usepackage{enumerate}
\usepackage[shortlabels]{enumitem}
\usepackage{setspace} 
\usepackage[]{layout}
\usepackage{color,soul} 
\usepackage{array}
\usepackage{pdflscape} 
\newcolumntype{L}[1]{>{\raggedright\let\newline\\\arraybackslash\hspace{0pt}}m{#1}}
\newcolumntype{C}[1]{>{\centering\let\newline\\\arraybackslash\hspace{0pt}}m{#1}}
\newcolumntype{R}[1]{>{\raggedleft\let\newline\\\arraybackslash\hspace{0pt}}m{#1}}
\usepackage{hyperref}

\pgfplotsset{compat=1.16} 

\usetikzlibrary{shapes,shapes.geometric,backgrounds}

\begin{document}

 \title{A Research Agenda on Pediatric Chest X-Ray:\\ Is Deep Learning Still in Childhood?}
\author{
\IEEEauthorblockN{
Afonso U. Fonseca\IEEEauthorrefmark{1}\textsuperscript{1},
Gabriel S. Vieira\IEEEauthorrefmark{2}\textsuperscript{2}, Fabrizzio Soares\IEEEauthorrefmark{3}\IEEEauthorrefmark{1}\textsuperscript{3},
Renato F. Bulcão-Neto\IEEEauthorrefmark{1}\textsuperscript{4},
}
\IEEEauthorblockA{\IEEEauthorrefmark{1}Federal University of Goiás, \textit{Pixelab Laboratory}. Goiânia - GO, Brazil\\
\IEEEauthorrefmark{2} Instituto Federal Goiano, \textit{Computer Vision Lab}, Uruta\'i/GO, Brazil\\
\IEEEauthorrefmark{3}Department of Computer Science, Southern Oregon University, Ashland/OR, USA\\
Email: \{\textsuperscript{1}afonsoueslei, \textsuperscript{4}rbulcao\}@ufg.br, \textsuperscript{2}gabriel.vieira@ifgoiano.edu.br, \textsuperscript{3}soaresf@sou.edu}
}

\maketitle
 
\definecolor{orange}{RGB}{220,57,18}
\definecolor{yellow}{RGB}{255,153,0}
\definecolor{blue}{RGB}{102,140,217}
\definecolor{green}{RGB}{16,150,24}
\definecolor{gray}{RGB}{220,220,220}

\makeatletter

\tikzstyle{chart}=[
    legend label/.style={anchor=west,align=left},
    legend box/.style={rectangle, draw=none, minimum size=10pt},
    axis/.style={black,semithick,->},
    axis label/.style={anchor=east,font={\tiny}},
]

\tikzstyle{bar chart}=[
    chart,
    bar width/.code={
        \pgfmathparse{##1/2}
        \global\let\bar@w\pgfmathresult
    },
    bar/.style={very thick, draw=white},
    bar label/.style={font={\bf\small},anchor=north},
    bar value/.style={font={\footnotesize}},
    bar width=.75,
]

\tikzstyle{pie chart}=[
    chart,
    slice/.style={line cap=round, line join=round, very thick,draw=white},
    pie title/.style={font={\tiny}},
    slice type/.style 2 args={
        ##1/.style={fill=##2},
        values of ##1/.style={}
    }
]

\pgfdeclarelayer{background}
\pgfdeclarelayer{foreground}
\pgfsetlayers{background,main,foreground}

\newcommand{\pie}[3][]{
    \begin{scope}[#1]
    \pgfmathsetmacro{\curA}{90}
    \pgfmathsetmacro{\r}{1}
    \def\c{(0,0)}
    \foreach \v/\s in{#3}{
        \pgfmathsetmacro{\deltaA}{\v/100*360}
        \pgfmathsetmacro{\nextA}{\curA + \deltaA}
        \pgfmathsetmacro{\midA}{(\curA+\nextA)/2}

        \path[slice,\s] \c
            -- +(\curA:\r)
            arc (\curA:\nextA:\r)
            -- cycle;
        \pgfmathsetmacro{\d}{max((\deltaA * -(.5/50) + 1) , .5)}

        \begin{pgfonlayer}{foreground}
        \path \c -- node[pos=\d,pie values,values of \s]{$\v\%$} +(\midA:\r);
        \end{pgfonlayer}

        \global\let\curA\nextA
    }
    \end{scope}
}

\newcommand{\legend}[2][]{
    \begin{scope}[#1]
    \path
        \foreach \n/\s in {#2}
            {
                ++(0,-10pt) node[\s,legend box] {} +(5pt,0) node[legend label] {\n}
            }
    ;
    \end{scope}
}


\begin{abstract}
Several reasons explain the significant role that chest X-rays play on supporting clinical analysis and early disease detection in pediatric patients, such as low cost, high resolution, low radiation levels, and high availability. In the last decade, Deep Learning (DL) has been given special attention from the computer-aided diagnosis research community, outperforming the state of the art of many techniques, including those applied to pediatric chest X-rays (PCXR). Due to this increasing interest, much high-quality secondary research has also arisen, overviewing machine learning and DL algorithms on medical imaging and PCXR, in particular. However, these secondary studies follow different guidelines, hampering their reproduction or improvement by third-parties regarding the identified trends and gaps. This paper proposes a ``deep radiography'' of primary research on DL techniques applied in PCXR images. We elaborated on a Systematic Literature Mapping (SLM) protocol, including automatic search on six sources for studies published from January 1, 2010, to May 20, 2020, and selection criteria utilized on a hundred research papers. As a result, this paper categorizes twenty-six relevant studies and provides a research agenda highlighting limitations, gaps, and trends for further investigations on DL usage in PCXR images. Besides the fact that there is no systematic mapping study on this research topic, to the best of authors' knowledge, this work organizes the process of finding and selecting relevant studies and data gathering and synthesis in a reproducible way.
\end{abstract}
\section{Introduction}
\label{sec:introduction}

Children under the age of fourteen represent at least $25\%$ of the world's population \cite{desa2019united}. Unfortunately, about $14\%$ of them suffer from respiratory diseases \cite{pearce2007worldwide} and $1.4$ millions of children die each year, mainly from pneumonia \cite{world2018top}. Many strategies have been applied to reduce these numbers, with particular attention to pediatric radiology supporting clinical analysis and early disease identification. 

Although it is not the most modern and accurate among clinical imaging assessment solutions, chest radiography (CXR) is a successful diagnostic imaging approach. CXR generates high-resolution images, requires few doses of ionizing radiation, is the lowest-cost exam compared to other medical imaging modalities, and is used in many countries due to its high availability and simplified acquisition mode \cite{zar2017advances}. 

CXR is usually used in the initial clinical tests before resorting to more expensive exams. Estimates point out that about $40\%$ of pediatric images are acquired as CXR, which corresponds to about $350$ million radiographs taken only from children every year \cite{garcia2011chest, blickman2009pediatric, xray2020}. 

However, pediatric imaging diagnosis is guided by advances in adult radiology whose protocols differ from pediatric X-rays. In this sense, technical challenges related to the environment, equipment for image acquisition, and a strategy that promotes the child's cooperation are rarely discussed. Besides, the high variation between children's anatomical structures and their disease patterns are neglected in the definition of more specific protocols \cite{menashe2016pediatric, arthur2000interpretation, thukral2015problems}.

A key challenge is how to make the diagnosis more accurate based on existing data. As the number of pediatric chest radiographic increases each year, intelligent solutions for medical decision-making are crucial to assist in disease prevention and treatment strategies and, consequently, to avoid the loss of children's lives. In this context, new cutting edge technologies are essential elements to meet this emerging demand, which directly affect the efficiency and cost of treatments \cite{zar2017advances} and have a positive impact on the lack of experienced radiologists around the world \cite{summers2018deep}.

Machine learning (ML) and computer vision have leveraged medical health, effectively assisting in diagnoses and interventions in small to high complexity problems. Deep learning (DL) is the dominant approach whose computational models have contributed to solving problems strictly limited by traditional approaches, including disease detection \cite{Salehinejad2019Cxr, Shin2016DcnnCAD}, classification \cite{Oh2020DLCovid19}, and segmentation \cite{Wang2018SegmentationDL, Menze2015BrainSegmentation}.

In radiology, ML algorithms have been widely used to recognize patterns and infer knowledge from a training data set. Due to the increasing number of primary research in ML applied to medical imaging along the last decades, it is natural that secondary research arises. For instance, Moore et al. \cite{moore2019machine} point out challenges of ML implementation in radiological images of children, and highlights the opaque character of convolution neural networks (CNNs). Litjens et al. \cite{litjens2017survey} present a broad review of the main DL concepts relevant to the analysis of medical images, including CXR. Ginneken \cite{van2017fifty} reviews techniques applied to chest images and shows that ML is the dominant technology in the detection of pulmonary nodules by computer-aided diagnosis (CAD) systems. Nonetheless, we have identified that the methodology adopted in these secondary studies follow different guidelines, hampering their reproduction or improvement by third-parties of the identified trends and gaps.



This paper describes a Systematic Literature Mapping (SLM) on applying deep learning techniques in pediatric chest X-ray images (PCXR). SLM gets a more coarse-grained overview of the topic of interest, providing categorization and visual maps of primary research results~\cite{kitchenham2011using,petersen2015guidelines}. Furthermore, the state of evidence in specific topics resulting from an SLM can be investigated further, e.g., through primary or secondary studies. In this light, our systematic mapping of the literature describes in detail the process of finding and selecting relevant studies, as well as data gathering and information analysis, in a way that the entire protocol can be reproduced later by other studies.


Based on guidelines for performing SLMs~\cite{kitchenham2011using,petersen2015guidelines}, we developed a protocol that describes our research questions, some strategies for identifying and selecting primary studies, and data analysis and synthesis activities. We performed an automatic search for studies published from January 1, 2010 to May 20, 2020, from six sources where we applied inclusion and exclusion criteria in one hundred research articles.

This original work categorizes twenty-six relevant studies on the metrics, datasets, neural network architectures, and imaging-related tasks addressed in each paper, among other aspects. As a result of our findings, we provide a research agenda highlighting limitations, gaps, and trends for further investigations on DL usage in PCXR images.

The remainder of this paper is organized as follows. Section \ref{sec:related-works} describes related work, and Section \ref{sec:sdudy-protocol} details the SLM protocol. Section \ref{sec:data-extraction} reports the data extraction process from primary studies, whereas Section \ref{sec:data-synthesis} presents a synthesis of our findings. Section \ref{sec:researchAgenda} introduces our research agenda, and Section \ref{sec:discussion} discusses the validity threats of this SLM and how we can work to mitigate them. Finally, Section \ref{sec:conclusion} suggests directions for future work.

\section{Related work}
\label{sec:related-works}

This section presents secondary studies focusing on deep learning and medical imaging.

Litjens et al. \cite{litjens2017survey} analyze the main DL concepts applied to the analysis of medical images and summarize more than three-hundred contributions, grouping them into image classification, object detection, segmentation, registration, and other tasks. Finally, they conclude that the DL will have a significant impact on the analysis of medical images as a whole.

Ginneken's work \cite{van2017fifty} describes the evolution of various techniques applied to chest imaging along fifty years. He discusses convolutional networks in CAD systems that simultaneously locate multiple different types of abnormalities on specific scans. Ginneken also concludes that DL will become the leading choice for image analysis.

Yasaka and Abe \cite{yasaka2018deep} review both DL and artificial intelligence in radiology, highlighting DL's contributions to detecting, diagnosing, staging, and sub classifying conditions in radiological images. Besides, they discuss limitations of DL, such as the low legibility and interpretation of the characteristics and calculations used in classification models. Consequently, this makes it difficult to resolve conflicts between trained models and the doctors' or the radiologists' judgments.

Tajbakhsh et al. \cite{Tajbakhsh2020embracing} review DL techniques applied explicitly to segmentation by raising issues related mainly to the scarcity and quality of the data set. In turn, Lee et al. \cite{lee2019deep} investigate the application of DL in CXR and CT images and note the DL's ability to handle new information and its high potential to overcome the limitations of existing CAD systems. They also point out that DL is still in its infancy in the medical field and that there is great concern with this technology in real-world clinical settings.


Although they are not systematic works, the literature reviews made by these authors are significant in helping other researchers to understand the current state of the art, limitations, challenges and future directions. What weighs at a disadvantage is that their results are not reproducible and are more subject to threats of validity, for example, bias in the selection and analysis of studies.

In the category of systematic literature studies, Pande et al. \cite{pande2016computer} performed a systematic review of computer-assisted detection approaches of pulmonary tuberculosis on digital CXR. As a result of an automatic search strategy over four sources, four-hundred-fifty-five papers published between January 1, 2010, to December 31, 2015, were retrieved. However, only five papers remained after their selection criteria. Despite this small number of studies, the authors found evidence, such as methodological limitations and generalization only for environments where pulmonary tuberculosis and HIV are less prevalent.


Compared to our work, Pande et al. \cite{pande2016computer} is a pioneering systematic study on DL techniques in medical imaging as well as our is the first, to the best of our knowledge, to prepare a systematic review on DL applied to pedriatric X-ray images. On the other hand, our work differs from others because our systematic mapping contributes to a research agenda that is based on a reproducible compilation of papers covering ten years of research, from 2010 to 2020.

\section{Systematic Mapping Protocol}\label{sec:sdudy-protocol}

A well-defined process guides every systematic literature mapping and comprises three
phases: planning, conducting, and publishing the results~\cite{kitchenham2011using,petersen2015guidelines}.


The planning phase starts with the SLM's objective statement and the definition of proper keywords, supported by a pilot search. Then, the protocol is created and evaluated to accommodate possible adjustments.

The conduction phase encompasses the activities of identification and selection of primary studies, data extraction, and synthesis. A search strategy and selection criteria allow the identification and selection of papers, respectively. Data extraction starts as soon as the relevant primary studies are selected, followed by a synthesis of these studies to answer the research questions of the mapping study.

Finally, synthesis results combine multiple formats, such as textual, tabular, and graphical descriptions, to be assessed by expert readers on the SLM's subject.

\begin{figure}[ht]
    \centering
    \includegraphics[width=\linewidth]{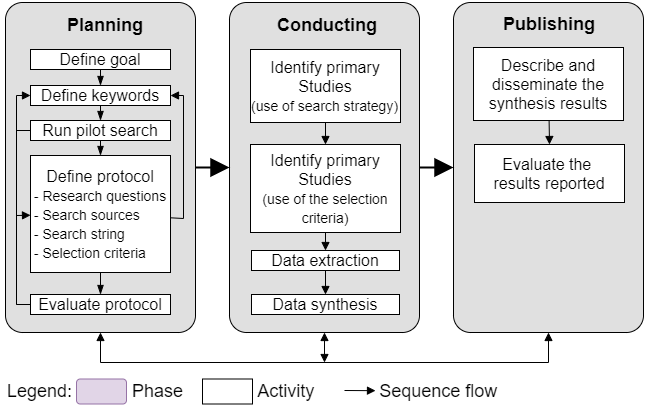}
    \caption{Systematic mapping protocol, adapted from \cite{kitchenham2011using,kudo2019requirement}.}
    \label{fig:phases}
\end{figure}

We used the StArt tool \cite{fabbri2016improvements} for the management of this whole process illustrated in Figure \ref{fig:phases}.


\subsection{Research Questions}\label{sub:research-questions}


We followed the widely-adopted criterion called PICO (\textbf{P}opulation, \textbf{I}ntervention, \textbf{C}omparison, \textbf{O}utcome) \cite{pai2004systematic} to help on the preparation of the research questions, as shown in Table \ref{tab:pico}.

\begin{table}[ht] \onehalfspacing
\small
    \centering
    \caption{PICO Analysis}
        \label{tab:pico}
    \begin{tabular}{m{0.3cm} m{7.7cm}} \hline
         P & Primary studies about DL techniques on PCXR images. \\ \hdashline[.2pt/.5pt]
         I & Tasks on PCXR images that use any DL solution.\\ \hdashline[.2pt/.5pt]
         C &  Not applicable in systematic literature mappings.\\ \hdashline[.2pt/.5pt]
         O &  Mapping of DL solutions applied to canonical tasks in PCXR images.\\ \hline
    \end{tabular}
\end{table}


The compilation of the primary and secondary research questions (RQs) and their respective justifications are presented as follows:

\begin{enumerate}[start=1,label={\empty}] 
    \item {\textbf{PRQ:} \textit{What is state of the art on DL and PCXR-related tasks?} To identify DL solutions' maturity level applied to canonical tasks in PCXR images and to investigate possible tasks not achieved by DL solutions. Examples of tasks include classification, detection, segmentation, registration, retrieval, imaging, enhancement, diagnostic, object recognition, localization, and prediction/prognostic.
    }
\end{enumerate}

To refine the primary research question (PRQ), we elaborated on further research questions (RQ) that this SLM's findings should answer:

\begin{enumerate}[start=1,label={\empty}] 
    \item{ \textbf{RQ1:} \textit{When and where were primary studies published?} To identify the venues of publication mostly targeted, the frequency of publications within a timeline, and support other RQs' answers.
    }
    
    \item { \textbf{RQ2:} \textit{Which tasks applied to PCXR images are most addressed by DL techniques?} Not only to map tasks applied to PCXR but also to identify likely gaps related to particular tasks.
    }
    
    \item{ \textbf{RQ3:} \textit{Which neural network architectures are most used?} To verify whether there is a dominant DL architecture over PCXR, following the classification of architectures proposed in~\cite{shresthaReviewDL2019}.
    }
    
    \item{ \textbf{RQ4:} \textit{Which metrics are used for assessment purposes?} To identify whether there is a prevalence usage of evaluation metrics regarding tasks.
    }
    
    \item{ \textbf{RQ5:} \textit{How is each DL technique in detail?} To describe the training, learning, and processing approaches in use and whether preprocessing steps are usual.
    }
    
    \item{ \textbf{RQ6:} \textit{What are the datasets used, and how are they organized?} To recognize the datasets available in the field, if they are public or private, and the number of images, the types of CXR, and additional information, such as reports, and other types of images.
    }
    
    \item{ \textbf{RQ7:} \textit{What is the primary type of contribution?} To specify each primary research's main contribution, such as algorithms, methods, models, metrics, system applications, frameworks, architectures, and processes.
    }

    \item{ \textbf{RQ8:} \textit{How can the research be classified?} According to the applied research method, to determine each study's research type, we follow the classification proposed in~\cite{petersen2015guidelines}.
    }

\end{enumerate}

\subsection{Pilot Search and Search String}

We carried out a pilot search with terms related to this SLM's theme to define a search string that balances coverage and precision. Moreover, the selection of these terms, i.e., keywords, synonyms,  and acronyms, counted on a DL expert's support. Examples of terms we used are as follows: deep learning, deep machine learning, deep inspection, artificial intelligence, artificial neural network, neural network, convolution network, convolutional neural network, CNN, recurrent neural network, RNN, deep belief network, DBN, autoencoder, chest, lung, breastplate, pulmonary, thoracic, X-ray, radiograph, radiogram, CXR,  child, pediatric, infant, baby, toddler, newborn, and neonate.

After examining each combination of terms in pilot searches through the Scopus database, we reached the following observations:

\begin{itemize}
    \item Regarding the synonyms for \textit{deep learning}, only {neural network, CNN}, and \textit{convolutional neural network} returned a high number of important papers, while the other terms did not represent any change.
    \item As synonym for the term \textit{chest}, only \textit{breastplate} did not return papers.
    \item As synonyms for \textit{pediatric} we discarded the terms, \textit{neonate} and  \textit{toddler}, due to the small number of papers returned.
\end{itemize}

Following that, we present the final search string as a result of the pilot search activity. Observe that the search string contains sub-strings referring to relevant terms for this SLM's context, such as ``thora'' and ``radiogra'' for thoracic, thorax, radiogram, and radiography, respectively.

\begin{center}
(\textit{deep learning} OR \textit{neural network} OR \textit{CNN} OR \textit{convolution net}) AND (((\textit{chest} OR \textit{lung} OR \textit{thora}) AND (\textit{x-ray} OR \textit{radiogra})) OR \textit{CXR}) AND (\textit{pediatr} OR \textit{paediatr} OR \textit{infant} OR \textit{baby} OR \textit{newborn} OR \textit{child})
\end{center}

\begin{figure*}
    \tikzstyle{block} = [rectangle, draw, text width=8em, text centered, rounded corners, node distance=2.6cm, minimum height=3.2em]
	\tikzstyle{line} = [draw, -latex]
	\tikzstyle{cloud} = [draw, ellipse,fill=white, node distance=1cm,
	minimum height=1em] 
	{\footnotesize
	\setlength{\fboxsep}{5pt}
	\fbox{
	\begin{tikzpicture}[node distance = 1cm, auto]
	\node [cylinder, shape border rotate=90, draw, minimum height=1.3cm, minimum width=1cm, shape aspect=.25] (DBs) {\textit{6 sources}} ;
	\node [block,right of = DBs, xshift=0.8cm,  yshift=-.75cm]  (AS)  {Automatic search};
	\node [label={[shift={(3.4,-1.75)}]178 identified}] {};
	\node [rotate=-12, label=below:] at (1.6,0.2) {Search string};
	\node [block,right of = AS,  xshift=0.5cm,  yshift=0cm]  (DSE) {Duplicate study exclusion};
	\node [label={[shift={(6.6,-1.75)}]78 duplicates}] {};
	\node [block,right of = DSE, xshift=0.5cm,  yshift=0cm]  (SS)  {Study selection};
	\node [rotate=30,label=below:] at (8.5,0) {100 selected};
	\node [label={[shift={(9.6,-1.75)}]74 excluded}] {};
	\node [block,right of = SS,  xshift=0.5cm,  yshift=0cm]  (PFE) {Preparation for extraction};
	\node [rotate=30, label=below:] at (11.6,0) {26 selected};
	\node [rotate=30, label=left:] at (14.8,0) {26 selected};
    \node [block,right of = PFE, xshift=0.5cm,  yshift=0cm]  (DEP){Data extraction and synthesis};
	
	\path [line] (AS)   -- (DBs);
	\path [line] (AS)   -- (DSE);
    \path [line] (DSE)  -- (SS);
    \path [line] (SS)   -- (PFE);
    \path [line] (PFE)  -- (DEP);
    
	\end{tikzpicture}
	}}
	\caption{An overview of the conduction phase and the activities of identification and selection of primary studies and data extraction and synthesis. The automatic search process retrieved 178 articles, with 78 duplicates removed. Then, we read the metadata of the articles remaining, upon which we applied the selection criteria. We selected 26 articles for full-text reading, none was discarded, and they contributed to answering our research questions. Figure adapted from \cite{kudo2019requirement}.}
	\label{fig:process-sms}
\end{figure*}
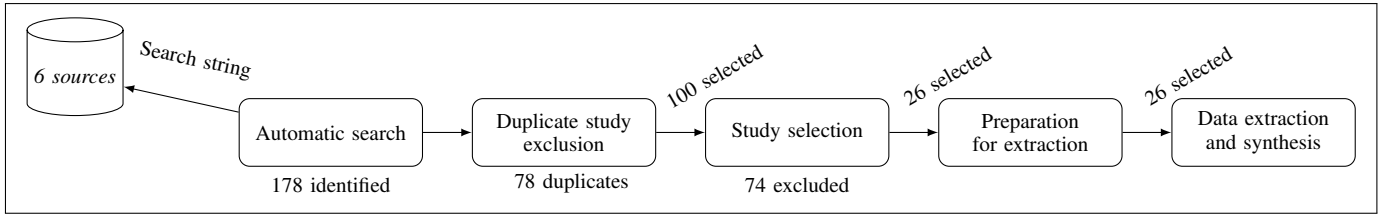

\subsection{Search Strategy}

Once we defined the search string, we chose a set of studies sources based on a list of prerequisites:

\begin{itemize}
    \item indexing sources of research on deep learning and medical imaging;
    \item web-based search mechanism with logical expressions support;
    \item search over studies title and abstract, at least; 
    \item exportation of search results for compatible formats with the StArt tool (e.g., BibTex, Medline, RIS).
\end{itemize}

As a result, we selected the following search engines and digital libraries: ACM Digital Library\footnote{Configured to \textit{The ACM Guide to Computing Literature} because it indexes a broader collection of papers.}, Engineering Village, Embase, IEEE Xplorer, PubMed, and Scopus.

We then adapted the final string to each search mechanism selected and searched for articles through their metadata contents (title, abstract, and keywords\footnote{Except in PubMed because it does not allow search in keywords.}). The final search strings generated for each source is found at \cite{fonseca2020pediatric}.

\subsection{Selection criteria}

We defined seven exclusion criteria (EC) to select studies both after the reading of metadata and full-text. The ECs are as follows:

\begin{description}
    \item[EC1.] The full text is not accessible.
    \item[EC2.] It is not an English-written paper.
    \item[EC3.] The paper is not published in journals or conference proceedings.
    \item[EC4.] The main subject of the paper is not DL applied to PCXR.
    \item[EC5.] The paper is published before 2010.
    \item[EC6.] It is not a primary research paper.
    \item[EC7.] It is an old version of a paper previously published.
\end{description}

The exclusion of a study occurs when it falls into at least one of such exclusion criterion. If not excluded, the study must meet the single inclusion criterion (\textbf{IC}):

\begin{description}
    \item[IC1.] An English-written paper describing original primary research on DL applied to PCXR, published after 2009 on a journal or conference proceedings.
\end{description}

\subsection{The Conducting Phase} \label{subsec:conducting}

Figure \ref{fig:process-sms} illustrates the whole process of the conduction phase, with the corresponding number of studies identified (178), duplicate (78), removed (74), selected (26), and analyzed (26). 

After removing duplicate entries from the same paper, the decision for rejecting a study or not through selection criteria relies on the reading of studies metadata. Besides, retrieved studies may be ignored while the full-text reading takes place. In this SLM, we excluded a seventy-four-studies group, only after reading their full texts. 


Table \ref{tab:breakdown-exclusion} presents a breakdown of the application of EC in each article database source. In particular, EC4 was responsible for the largest number of excluded papers, probably because low precision rates are commonplace in systematic literature mappings \cite{petersen2015guidelines,kudo2019requirement}.

\begin{table}[ht] 
    \onehalfspacing
    \setlength{\tabcolsep}{5pt}
    \centering
    \caption{Breakdown of the exclusion criteria per source.}
    \resizebox{\columnwidth}{!}{
    \begin{tabular}{m{2cm}cccccccc} \hline
         \textbf{}  & \textbf{EC1} & \textbf{EC2} & \textbf{EC3} & \textbf{EC4} & \textbf{EC5} & \textbf{EC6} & \textbf{EC7} & \textbf{TOTAL}\\ \hline
         ACM DL              & 0   &  0  &  0  & 1   & 0   & 0   & 0 &  \textbf{1}\\ 
         Eng. Village        & 1   &  1  &  0  & 0   & 1   & 0   & 0 &  \textbf{3}\\
         Embase              & 1   &  0  &  4  & 10  & 0   & 0   & 0 &  \textbf{14}\\
         IEEE Xplorer        & 0   &  0  &  0  & 3   & 1   & 0   & 0 &  \textbf{4}\\ 
         PubMed              & 0   &  0  &  0  & 0   & 0   & 0   & 0 &  \textbf{0}\\ 
         Scopus              & 5   &  1  &  0  & 26  & 10  & 8   & 1 &  \textbf{51}\\ \hline 
         \textbf{TOTAL}               & \textbf{7}   &  \textbf{2}  &  \textbf{4}  & \textbf{40}  & \textbf{12}  & \textbf{8}   & \textbf{1} &  \textbf{74}\\ \hline
    \end{tabular}}
    \label{tab:breakdown-exclusion}
\end{table}

After applying the selection criteria, twenty-six studies remained for the extraction process, as shown in Table \ref{tab:selected-ps}. Throughout this paper, we identify each relevant study as Sn.


Table \ref{tab:duplicateStudies} shows these same twenty-six studies and their respective source as well, including duplicates. The $\circ$ symbol represents each study instance excluded because of copies in more than one source. The $\bullet$ symbol, in turn, represents the instance of a study available for the extraction activity.

Observe that only six studies had no duplicates (S2, S4, S11, S16, S22, and S23). Furthermore, Scopus and IEEE together index the twenty-six relevant studies of this SLM: one study is exclusively indexed by IEEE (S16), and five ones by Scopus (S2, S4, S11, S22, and S23). Besides, Scopus retrieves more than 92\% of the relevant selected studies of this SLM (24 of 26). These findings may help in future systematic literature studies on deep learning and PCXR. 

\begin{table*}
    \centering
    \onehalfspacing
    \caption{Selected primary studies}
    \resizebox{\textwidth}{!}{%
    \begin{tabular}{cm{16.5cm}ccc} \hline
        \textbf{ID} & \textbf{TITLE} & \textbf{REF.}  \\ \hline
\textbf{S1} & A generic approach to lung field segmentation from chest radiographs using deep space and shape learning  & \cite{Mansoor20201206} \\ \hdashline[.2pt/.5pt]
\textbf{S2} & A novel transfer learning based approach for pneumonia detection in chest X-ray images                    & \cite{Chouhan2020} \\ \hdashline[.2pt/.5pt]
\textbf{S3} & A transfer learning method with deep residual network for pediatric pneumonia diagnosis                   & \cite{Liang2020} \\ \hdashline[.2pt/.5pt]
\textbf{S4} & An efficient deep learning approach to pneumonia classification in healthcare                             & \cite{Stephen2019} \\ \hdashline[.2pt/.5pt]
\textbf{S5} & Automated deep learning design for medical image classification by health-care professionals with no coding experience: a feasibility study & \cite{Faes2019e232} \\ \hdashline[.2pt/.5pt]    
\textbf{S6} & Automated pneumonia diagnosis using a customized sequential convolutional neural network                  & \cite{Siddiqi201964} \\ \hdashline[.2pt/.5pt]
\textbf{S7} & Automatic catheter and tube detection in pediatric x-ray images using a scale-recurrent network and synthetic data & \cite{Yi2020181} \\ \hdashline[.2pt/.5pt]          
\textbf{S8} & Automatic tissue characterization of air trapping in chest radiographs using deep neural networks         & \cite{Mansoor201697} \\ \hdashline[.2pt/.5pt]
\textbf{S9} & Classification of bacterial and viral childhood pneumonia using deep learning in chest radiography        & \cite{Gu201888} \\ \hdashline[.2pt/.5pt]
\textbf{S10} & Classification of images of childhood pneumonia using convolutional neural networks                      & \cite{Saraiva2019112} \\ \hdashline[.2pt/.5pt]
\textbf{S11} & Classification of pneumonia from x-ray images using siamese convolutional network                        & \cite{Prayogo20201302} \\ \hdashline[.2pt/.5pt]
\textbf{S12} & Deep learning method for automated classification of anteroposterior and posteroanterior chest radiographs    & \cite{Kim2019925} \\ \hdashline[.2pt/.5pt]   
\textbf{S13} & Deep learning to automate Brasfield chest radiographic scoring for cystic fibrosis                       & \cite{Zucker2020131} \\  \hdashline[.2pt/.5pt]   
\textbf{S14} & Deep learning, reusable and problem-based architectures for detection of consolidation on chest X-ray images   & \cite{Behzadi-khormouji2020} \\ \hdashline[.2pt/.5pt]     
\textbf{S15} & Detecting pneumonia in chest radiographs using convolutional neural networks                             & \cite{Ureta2020} \\ \hdashline[.2pt/.5pt]
\textbf{S16} & Detection of pediatric pneumonia from chest x-Ray images using CNN and transfer learning                 & \cite{9091755} \\ \hdashline[.2pt/.5pt]
\textbf{S17} & Discriminant analysis deep neural networks                                                               & \cite{Li2019} \\ \hdashline[.2pt/.5pt]
\textbf{S18} & Identifying medical diagnoses and treatable diseases by image-based deep learning                        & \cite{Kermany2018-Guangzhou} \\ \hdashline[.2pt/.5pt]
\textbf{S19} & Learning to recognize chest-xray images faster and more efficiently based on multi-kernel depthwise convolution & \cite{Hu202037265} \\ \hdashline[.2pt/.5pt]        
\textbf{S20} & LungAIR: An automated technique to predict hospitalization due to LRTI using fused information           & \cite{Mansoor2018} \\ \hdashline[.2pt/.5pt] 
\textbf{S21} & Marginal shape deep learning: Applications to pediatric lung field segmentation                          & \cite{Mansoor2017} \\ \hdashline[.2pt/.5pt]
\textbf{S22} & Pulmonary rontgen classification to detect pneumonia disease using convolutional neural networks         & \cite{Rustam20201522} \\ \hdashline[.2pt/.5pt]
\textbf{S23} & Simultaneous lung field detection and segmentation for pediatric chest radiographs                       & \cite{Zhang2019594} \\ \hdashline[.2pt/.5pt]
\textbf{S24} & Two-stage deep learning architecture for pneumonia detection and its diagnosis in chest radiographs      & \cite{Narayanan2020} \\ \hdashline[.2pt/.5pt]
\textbf{S25} & Using deep-learning techniques for pulmonary-thoracic segmentations and improvement of pneumonia diagnosis in pediatric chest radiographs    & \cite{Longjiang20191617} \\ \hdashline[.2pt/.5pt]     
\textbf{S26} & Visualizing and explaining deep learning predictions for pneumonia detection in pediatric chest radiographs & \cite{Rajaraman2019} \\   \hline
    \end{tabular}}
    \label{tab:selected-ps}
\end{table*}


\begin{table*} 
\doublespacing
\centering
\setlength{\tabcolsep}{4pt}
\caption{Selected papers and digital libraries}
{
\resizebox{\linewidth}{!}{%
\begin{tabular}{m{2cm}cccccccccccccccccccccccccc} \hline

\textbf{STUDY} & \textbf{S1} & \textbf{S2} & \textbf{S3} & \textbf{S4} & \textbf{S5} & \textbf{S6} & \textbf{S7} & \textbf{S8} & \textbf{S9} & \textbf{S10} & \textbf{S11} & \textbf{S12} & \textbf{S13} & \textbf{S14} & \textbf{S15} & \textbf{S16} & \textbf{S17} & \textbf{S18} & \textbf{S19} & \textbf{S20} & \textbf{S21} & \textbf{S22} & \textbf{S23} & \textbf{S24} & \textbf{S25} & \textbf{S26}\\ \hline

ACM DL                  &   &   &   &   &   & $\circ$ &   &   & $\circ$ &   &   &   &   &   &   &   &   &   &   &   &   &   &   &   &   &   \\ \hdashline[0.2pt/.5pt]
Eng. Village  & $\circ$ &   & $\circ$ &   &   & $\circ$ & $\circ$ & $\circ$ & $\circ$ & $\circ$ &   & $\circ$ &   & $\circ$ & $\circ$ &   & $\circ$ &   & $\circ$ & $\circ$ & $\circ$ &   &   & $\circ$ &   & $\circ$ \\ \hdashline[0.2pt/.5pt]
Embase               & $\circ$ &   & $\circ$ &   & $\circ$ &   & $\circ$ & $\circ$ &   &   &   & $\circ$ & $\circ$ & $\circ$ &   &   &   & $\circ$ &   &   &   &   &   &   & $\circ$ &   \\ \hdashline[0.2pt/.5pt]
IEEE Xplorer         & $\bullet$ &   &   &   &   &   &   & $\circ$ &   &   &   &   &   &   &   & $\bullet$ & $\circ$ &   &   &   &   &   &   &   &   &   \\ \hdashline[0.2pt/.5pt]
PubMed               & $\circ$ &   & $\circ$ &   &   &   & $\circ$ &   &   &   &   & $\circ$ & $\circ$ & $\circ$ &   &   &   & $\circ$ &   &   & $\circ$ &   &   &   & $\circ$ &   \\ \hdashline[0.2pt/.5pt]
Scopus               &   & $\bullet$ & $\bullet$ & $\bullet$ & $\bullet$ & $\bullet$ & $\bullet$ & $\bullet$ & $\bullet$ & $\bullet$ & $\bullet$ & $\bullet$ & $\bullet$ & $\bullet$ & $\bullet$ &   & $\bullet$ & $\bullet$ & $\bullet$ & $\bullet$ & $\bullet$ & $\bullet$ & $\bullet$ & $\bullet$ & $\bullet$ & $\bullet$ \\ \hline

\end{tabular}}}
\label{tab:duplicateStudies}
\end{table*}


\begin{table*}[]
\onehalfspacing
\setlength{\tabcolsep}{4pt}
\caption{Publication venues and the corresponding ranking indexes.}
\resizebox{\textwidth}{!}{%
\begin{tabular}{L{12cm}C{1.3cm}C{.9cm}C{.9cm}C{1.45cm}C{1.7cm}L{1.55cm}} \hline
\textbf{EVENTS} & \textbf{H-index} & \textbf{H5} & \textbf{IF} & \textbf{SJR 2019} & \textbf{YEAR} & \textbf{STUDIES} \\ \hline
International Conference of the IEEE Engineering in Medicine and Biology Society & 38 & 39 & N/A & N/A & 2016 & S8 \\ \hdashline[.2pt/.5pt]
International Conference on Biomedical Engineering Systems and Technologies & N/A & 6 & N/A & 0.12 & 2019 & S10 \\ \hdashline[.2pt/.5pt]
International Conference on Deep Learning Technologies & N/A & 10 & N/A & N/A & 2019 & S6 \\ \hdashline[.2pt/.5pt]
International Conference on Emerging Technologies in Computer Engineering: Machine Learning and Internet of Things & N/A & N/A & N/A & N/A & 2020 & S16 \\ \hdashline[.2pt/.5pt]
International Conference on Information Sciences and Systems & N/A & 22 & N/A & N/A & 2019 & S17 \\ \hdashline[.2pt/.5pt]
International Conference on Machine Vision &  N/A & 15 & N/A & N/A & 2019 & S15 \\ \hdashline[.2pt/.5pt]
International Conference on Medical Image Computing and Computer-Assisted Intervention & 14 & 41 & N/A & N/A & 2019 & S23 \\ \hdashline[.2pt/.5pt]
International Conference on Multimedia and Image Processing & N/A & N/A & N/A & N/A  & 2018 & S9 \\ \hdashline[.2pt/.5pt]
International Symposium on Medical Information Processing and Analysis &  N/A &  9 & N/A & N/A & 2018 & S20 \\ \hdashline[.2pt/.5pt]
SPIE Medical Imaging & 50 & N/A &  N/A & 0.27 & 2017, 2020, 2019 & S21, S24, S26 \\ \hline

\textbf{JOURNALS}& \textbf{H-index} & \textbf{H5} & \textbf{IF} & \textbf{SJR 2019} & \textbf{YEAR} & \textbf{STUDIES}  \\ \hline
Applied Sciences & 35 & 2 & 2.49 & 0.42 & 2020 &  S2 \\ \hdashline[.2pt/.5pt]
Cell & 747 & 269 & 27.35 & 24.7 & 2018 & S18 \\ \hdashline[.2pt/.5pt]
Computer Methods and Programs in Biomedicine & 92 & 55 & 4.26 & 0.95 & 2019, 2020 & S3, S14 \\ \hdashline[.2pt/.5pt]
IEEE Access & 86 & 119 & 4.64 & 0.78 & 2020 & S19 \\ \hdashline[.2pt/.5pt]
IEEE Transactions on Biomedical Engineering  & 185 & 11 & 4.78 & 1.41 & 2019 & S1 \\ \hdashline[.2pt/.5pt]
Journal of Cystic Fibrosis & 71 & 39 & 3.51 & 1.41 & 2020  & S13 \\ \hdashline[.2pt/.5pt]
Journal of Digital Imaging & 51 & 34 & 2.99 & 0.97 & 2019 & S7, S12 \\ \hdashline[.2pt/.5pt]
Journal of Healthcare Engineering & 23 & 26 & 1.51 & 0.42 & 2019 & S4 \\ \hdashline[.2pt/.5pt]
Pediatric Pulmonology & 102 & 38 & 2.64 & 0.93 & 2019 & S25 \\ \hdashline[.2pt/.5pt]
Telkomnika & 18 & 18 & N/A & 0.21 & 2020 & S11, S22 \\ \hdashline[.2pt/.5pt]
The Lancet Digital Health & 4 & N/A & N/A & N/A & 2019 & S5 \\ \hline
\\
\end{tabular}}
\label{tab:publicVenues&IndexRanking}
\small{  \textbf{Notes:} 
H-index and SRJ (2019) search from Scimago Institutions Rankings in the url: \textcolor{blue}{\url{https://www.scimagojr.com}}, IF search from Resurchify website in the url: \textcolor{blue}{\url{https://www.resurchify.com/impact-factor-search.php}} and current H5 obtained on the Google Scholar website in the url: \textcolor{blue}{\url{https://scholar.google.com}}. All indexes are up to date 01/09/2020 date of consultation and \textbf{N/A} indicates index is not available.}
\end{table*}

\section{Data Extraction}\label{sec:data-extraction}

This section describes the most important aspects and information extracted from the full-text reading of the 26 primary studies selected which include:

\begin{itemize}
    \item the main objective and respective RQs (research questions);
    \item the selection methods of primary studies; 
    \item the evidence collected from the synthesis of these studies. 
\end{itemize}


\begin{table*}
\setlength{\tabcolsep}{4pt}
\centering
\onehalfspacing
\caption{DL architectures adopted by task in the studies}
    \resizebox{\textwidth}{!}{%
\begin{tabular}{lllc} \hline

\textbf{TASKS} & \textbf{ARCHITECTURES} & \textbf{STUDIES} & \textbf{TOTAL}\\ \hline
Classification & Autoencoder, CNN, Residual Network & S2, S4-S6, S8-S12, S14, S15, S17-S19, S22, S24-S26 & \textbf{18}\\\hdashline[.2pt/.5pt]
Detection & CNN, Recurrent Neural Network & S2, S6, S7, S14, S15, S23, S26 & \textbf{7}\\ \hdashline[.2pt/.5pt]
Diagnostic & CNN, Residual Network & S3, S5, S16, S18, S19 & \textbf{5}\\ \hdashline[.2pt/.5pt]
Segmentation & Autoencoder, CNN, Residual Network & S1, S9, S21, S23, S25 & \textbf{5}\\ \hdashline[.2pt/.5pt]
Prediction/Prognostic & Autoencoder, Residual Network & S13, S20 & \textbf{2}\\ \hline 
\end{tabular}}
\label{tab:DLarchitecturesVsTasks}
\end{table*}

Two independent reviewers extracted the data using a standardized form in the extraction phase, while a third, more experienced reviewer was left to resolve doubts and disagreements. The form consists of the following data fields: year of publication, publishing vehicle, task covered, DL architecture, metrics, learning, and processing approaches, preprocessing steps, use of international standards, use of public dataset (and type of), amount of PCXR images used, main contribution, research type, and research method.

A note is that some features may appear in several studies; therefore, the totals may not always correspond to 100$\%$. 

For a more complete version of this standard question form and its relationship to RQs, the former being normally used to answer the latter, see \cite{fonseca2020pediatric}.

It is worth mentioning that all 26 primary studies selected for data extraction and synthesis were retrieved from only 2 of the 6 researched digital libraries, Scopus and IEEE Xplorer (see Table \ref{tab:duplicateStudies}), through automatic search and according to the criteria adopted in their selection. Table \ref{tab:publicVenues&IndexRanking} presents a list with the places of publication and respective indexes of classification of these studies. Besides, we can see that all date from the last five years are distributed among events and journals in Table \ref{tab:StudiesByYear&Vehicle}. 

\begin{table}[ht]
\centering
\onehalfspacing
\caption{Studies per year and publication venue.}
\resizebox{.46\textwidth}{!}{
\begin{tabular}{lC{2.5cm}C{2.5cm}c} \hline
\multicolumn{1}{c}{\multirow{2}{*}{\textbf{YEAR}}} & \multicolumn{2}{c}{\textbf{PUBLICATION VENUE}} & \multirow{2}{*}{\textbf{TOTAL}} \\ \cline{2-3}
 & \multicolumn{1}{c}{\textbf{EVENT}} & \multicolumn{1}{c}{\textbf{JOURNAL}} &  \\ \hline
2016 & S8 &  &   \textbf{1} \\ \hdashline[.2pt/.5pt]
2017 & S21 &  &   \textbf{1} \\ \hdashline[.2pt/.5pt]
2018 & S9, S20 &  S18 & \textbf{3} \\ \hdashline[.2pt/.5pt]
2019 & S6, S10, S15, S17, S23, S26 & S1, S3, S4, S5, S7, S12, S25 & \textbf{13} \\ \hdashline[.2pt/.5pt]
2020 & S16, S24 & S2, S11, S13, S14, S19, S22 &  \textbf{8} \\ \hline
\textbf{Total} & \multicolumn{1}{c}{\textbf{12}} & \multicolumn{1}{c}{\textbf{14}} &  \multicolumn{1}{c}{\textbf{26}} \\ \hline
\end{tabular}}
\label{tab:StudiesByYear&Vehicle}
\end{table}


\begin{table}
\setlength{\tabcolsep}{6pt}
\caption{Metrics used in each study.}
\resizebox{0.49\textwidth}{!}{
\begin{tabular}{L{2.8cm}L{4cm}c} \hline
\textbf{METRIC} & \textbf{STUDIES} & \textbf{TOTAL} \\ \hline
Recall & S2, S3, S5-S12, S14-S19, S22, S26 & \textbf{17} \\ \hdashline[.2pt/.5pt]
Precision & S2, S3, S5-S7, S9-S12, S14-S19, S26 & \textbf{16} \\ \hdashline[.2pt/.5pt]
Accuracy & S2, S4, S6, S9-S12, S14-S19, S22, S24, S26 & \textbf{16} \\ \hdashline[.2pt/.5pt]
Area under the ROC Curve (AUC) & S2, S3, S5, S9, S10, S12, S14, S18, S19, S24-S26 & \textbf{12} \\ \hdashline[.2pt/.5pt]
*Other metrics & S1, S7, S13, S19-S21, S23, S25, S26 & \textbf{9} \\ \hdashline[.2pt/.5pt]
Dice coefficient & S1, S8, S9, S21, S23, S25 & \textbf{6} \\ \hdashline[.2pt/.5pt]
F1 Score & S3, S8, S11, S15, S16, S26 & \textbf{6} \\ \hdashline[.2pt/.5pt]
Confusion matrix & S5, S6, S16, S24, S26 & \textbf{5} \\ \hdashline[.2pt/.5pt]
Loss & S4, S11, S16, S22 & \textbf{4} \\ \hdashline[.2pt/.5pt]
Average contour distance (ACD) & S1, S21, S23 & \textbf{3} \\ \hdashline[.2pt/.5pt]
Speed & S19, S21 & \textbf{2} \\ \hline
\end{tabular}}
\label{tab:metricsUsed}

\small{\textbf{*} Other metrics include Spearman's correlation analysis, Scoring risk and Matthews correlation coefficient (MCC) just to name a few.}
\end{table}


\begin{table}[!hb]
\onehalfspacing
\setlength{\tabcolsep}{5pt}
\caption{Preprocessing (PP) vs Computation type}
\resizebox{0.49\textwidth}{!}{
\begin{tabular}{clL{3.8cm}c} \hline
\multicolumn{1}{c}{\textbf{PP}} & \multicolumn{1}{c}{\textbf{COMPUTING}} & \textbf{STUDIES} & \textbf{TOTAL} \\ \hline

\multirow{3}{*}{Yes} & Sequential & S1, S6, S8, S24 & \textbf{4}\\ 
 & Parallel &  S2-S4, S7, S9, S12-S16, S25, S26 & \textbf{12}\\ 
 & Not reported & S10, S11, S20, S21, S22 & \textbf{5} \\ \hdashline[.2pt/.5pt]
 \multicolumn{2}{l}{\textbf{Total PP}} &  & \textbf{21} \\ \hline
\multirow{3}{*}{No} & Sequential & - & - \\ 
 & Parallel & S5, S18, S19, S23 & \textbf{4} \\ 
 & Not reported &  S17 & \textbf{1}\\ \hdashline[.2pt/.5pt]
 \multicolumn{2}{l}{\textbf{Total No-PP}} &  & \textbf{5} \\ \hline
\end{tabular}}
\label{tab:PreprocessingVsComputationType}
\end{table}

The data extracted from these studies reveal important information for this SLM study, such as consensual definitions, common practices, some current panoramas and possible trends, while they are fundamental for us to be able to answer our research questions and propose the research agenda.



In the tasks addressed in the studies and in the adopted DL architectures (Table \ref{tab:DLarchitecturesVsTasks}), we can see that a same architecture was used in different tasks. Likewise, a same task was investigated by different architectures. Also, a research study can cover more than one task. For example, in the \textbf{S2} study not only the detection task was investigated but also the classification.

Table \ref{tab:metricsUsed} presents a list of metrics used in the selected studies. Metrics like Recall, Precision and Accuracy are widely used by several studies while other metrics like DICE coefficient and Average Contour Distance (ACD) are more associated with studies that deal with the segmentation task.


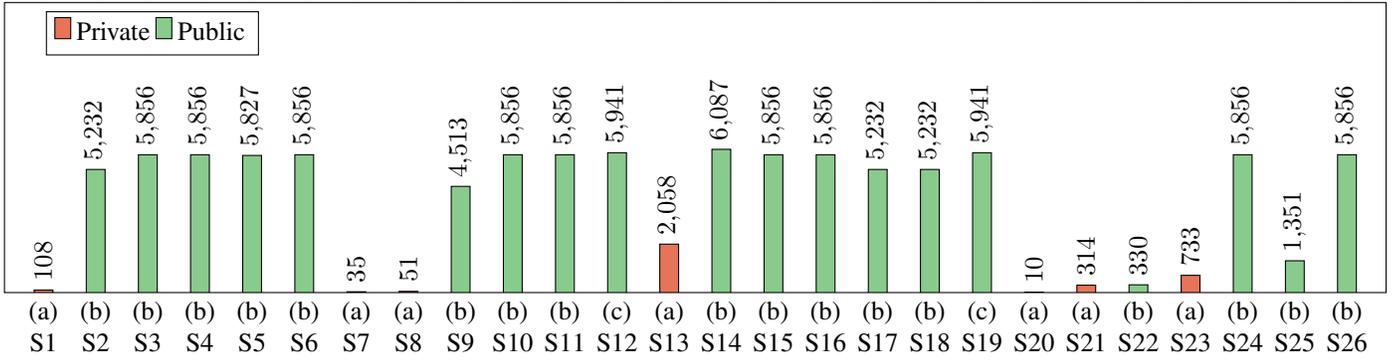
\begin{figure*}
    \centering
    \begin{tikzpicture}
    \begin{axis}[
        xticklabel style={rotate=0, align=center},
        legend columns=2,legend pos=north west, 
        major tick length=0cm,
        table/col sep      = semicolon,
        xtick=data,
        ymin=0,  ymax=8500, bar width=2.5mm, 
        enlarge y limits={upper,value=0.45},
        enlarge x limits={0.03}, 
        nodes near coords, width = 1.1\textwidth,
        height=.3\textwidth,
        ybar stacked, ytick=\empty,
        every node near coord/.append style={rotate=90, anchor=west},
        symbolic x coords={(a)\\S1,(b)\\S2,(b)\\S3,(b)\\S4,(b)\\S5,(b)\\S6,(a)\\S7,(a)\\S8,(b)\\S9,(b)\\S10,(b)\\S11,(c)\\S12,(a)\\S13,(b)\\S14,(b)\\S15,(b)\\S16,(b)\\S17,(b)\\S18,(c)\\S19,(a)\\S20,(a)\\S21,(b)\\S22,(a)\\S23,(b)\\S24,(b)\\S25,(b)\\S26},
        ]
        \addplot [style={fill=orange!70}] coordinates { ({(a)\\S1},108) ({(b)\\S2},0) ({(b)\\S3},0) ({(b)\\S4},0) ({(b)\\S5},0) ({(b)\\S6},0) ({(a)\\S7},35) ({(a)\\S8},51) ({(b)\\S9},0) ({(b)\\S10},0) ({(b)\\S11},0) ({(c)\\S12},0) ({(a)\\S13},2058) ({(b)\\S14},0) ({(b)\\S15},0) ({(b)\\S16},0) ({(b)\\S17},0) ({(b)\\S18},0) ({(c)\\S19},0) ({(a)\\S20},10) ({(a)\\S21},314) ({(b)\\S22},0) ({(a)\\S23},733) ({(b)\\S24},0) ({(b)\\S25},0) ({(b)\\S26},0) };
        \addplot [style={fill=green!50}] coordinates { ({(a)\\S1},0) ({(b)\\S2},5232) ({(b)\\S3},5856) ({(b)\\S4},5856) ({(b)\\S5},5827) ({(b)\\S6},5856) ({(a)\\S7},0) ({(a)\\S8},0) ({(b)\\S9},4513) ({(b)\\S10},5856) ({(b)\\S11},5856) ({(c)\\S12},5941) ({(a)\\S13},0) ({(b)\\S14},6087) ({(b)\\S15},5856) ({(b)\\S16},5856) ({(b)\\S17},5232) ({(b)\\S18},5232) ({(c)\\S19},5941) ({(a)\\S20},0) ({(a)\\S21},0) ({(b)\\S22},330) ({(a)\\S23},0) ({(b)\\S24},5856) ({(b)\\S25},1351) ({(b)\\S26},5856) };
    \legend{Private,Public}
  \end{axis}
  \end{tikzpicture}
  \caption{Amount of PCXR images used per study from (a) private datasets, (b) Guangzhou Women and Children's Medical Center \cite{Kermany2018-Guangzhou}, and (c) National Institutes of Health (NIH) Clinical Center \cite{wang2017-NIH-dataset}. Some studies have used non-PCXR images, but those images are not counted here.}
  \label{fig:graf-amount-datasets}
\end{figure*}


\begin{table}[]
\onehalfspacing
\centering
\caption{Results proposed by the studies}
\begin{tabular}{m{2.5cm}m{3.5cm}c} \hline
\multicolumn{1}{m{2.5cm}}{\textbf{RESULT TYPE}} & \multicolumn{1}{l}{\textbf{STUDIES}} & \multicolumn{1}{c}{\textbf{TOTAL}} \\ \hline
Algorithm & S1, S7, S10-S12, S14-17, S20, S22, S24 & \textbf{12} \\ \hdashline[.2pt/.5pt]
Framework & S1-S4, S6, S8-S10, S18, S19, S21, S23, S25, S26 & \textbf{14} \\ \hdashline[.2pt/.5pt]
Others & S5, S13 & \textbf{2} \\ \hline
\end{tabular}
\label{tab:resultsProposedByStudies}
\end{table}


\begin{table}[ht]
\setlength{\tabcolsep}{4pt}
\onehalfspacing
\centering
\caption{Research type and method of studies}
\begin{tabular}{L{1.8cm}C{1.8cm}C{2.5cm}c}
\hline
\multicolumn{1}{C{1.8cm}}{\textbf{METHOD}} & \multicolumn{1}{C{1.8cm}}{\textbf{TYPE}} & \textbf{STUDY(IES)} & \textbf{TOTAL} \\ \hline
\multirow{2}{*}{\begin{tabular}[l]{@{}l@{}}Controlled \\ Experiments\end{tabular}} & Evaluation & S1-S4, S6-S12, S14-S26 & \textbf{24} \\  \cdashline{2-4}[.2pt/.5pt]
 & Validation & S5, S13 & \textbf{2} \\ \hline
\multirow{2}{*}{Case Study} & Evaluation & - & \textbf{-} \\  \cdashline{2-4}[.2pt/.5pt]
 & Validation & S14 & \textbf{1} \\ \hline
\end{tabular}
\label{tab:ResearchType&Method}
\end{table}



Concerning to CXR images, almost always taken in frontal projection, anteroposterior (AP) or posteroanterior (PA), the use of public data sets has been the choose predominant in most studies. However, for the pediatric chest X-ray images, the number of public data sets is still small. In the selected studies only the Guangzhou data set \cite{Kermany2018-Guangzhou} and a subset of NIH data set \cite{wang2017-NIH-dataset} present PCXR images (details in Figure \ref{fig:graf-amount-datasets}).

Regarding common practices, the 26 studies reveal the massive use of preprocessing approaches and parallel computing as shown in Table  \ref{tab:PreprocessingVsComputationType} while CNN architectures showed dominance for various tasks and supervised learning was absolute in all studies. On the other hand, no international standards or guidelines were mentioned as the basis for specific or general tasks regarding the use of DL in PCXR images.



Table \ref{tab:resultsProposedByStudies} shows that the selected studies provided an algorithm or a framework as a result. However, they did not mention a finished product or tool for registration or use in clinical practice, but they emphasize the need for more testing and studies for future adoption. We believe that the next steps for studies like these will involve general and intelligible solutions that may be easy to check, as well as more in line with ethical dilemmas.

Finally, our data extraction stage reveals, as presented in Table \ref{tab:ResearchType&Method}, that the selected studies mostly follow in the field of evaluation-type research, adopting controlled experiments as their research method.

\vspace{.3cm}

\section{Data Synthesis}\label{sec:data-synthesis}

The answers shown next represent the synthesis of the results obtained in the extraction step. Far from being conclusive answers, they are much more of a strong indication of the possible paths taken in the research topic under analysis.
\vspace{.3cm}

\begin{figure}[hb]
    \centering
    \begin{tikzpicture}
    \begin{axis}[
        ybar stacked, ymin=0,  ymax=12.1, ytick = \empty, bar width=8mm,
        axis x line* = bottom, axis y line* = right,
        xtick=data, nodes near coords, width= 0.56\textwidth, height = 0.35\textwidth,
        ymajorgrids = true, legend pos=north west, xticklabels, symbolic x coords={2016,2017,2018,2019,2020},
    ]
    \addplot[style={black,fill=blue!10}] coordinates { ({2016},0) ({2017},1) ({2018},1) ({2019},7) ({2020},6) };
    \addplot[style={black,fill=blue!80}] coordinates { ({2016},1) ({2017},0) ({2018},2) ({2019},5) ({2020},3) };
    \legend{Periodicals,Events}
  \end{axis}
  \end{tikzpicture}
  \caption{Studies for extraction by publication vehicle per year.}
  \label{fig:grafYearVsVehicle}
\end{figure}
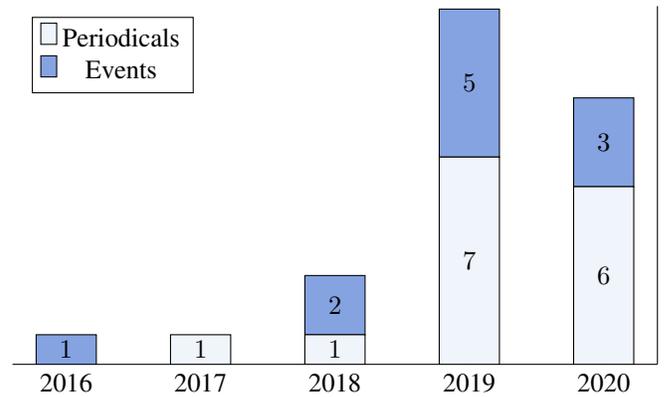

\textbf{RQ1:} \textit{When and where were primary studies published?} 
In view of what has observed in Tables \ref{tab:publicVenues&IndexRanking} and \ref{tab:StudiesByYear&Vehicle} and now in Figure \ref{fig:grafYearVsVehicle}, we can say that this research agenda deals with a very hot, current and highly relevant topic, which has aroused growing interest on the part of the scientific community of researchers. The studies date from all 5 years and are distributed in a large offer of publications of events and journals without vehicle specific prevalence so far and with mostly significant ranking indexes, only for two studies all of events (S9, S16) no index was found.


\begin{figure*}
    \centering
    \includegraphics[width=.99\textwidth]{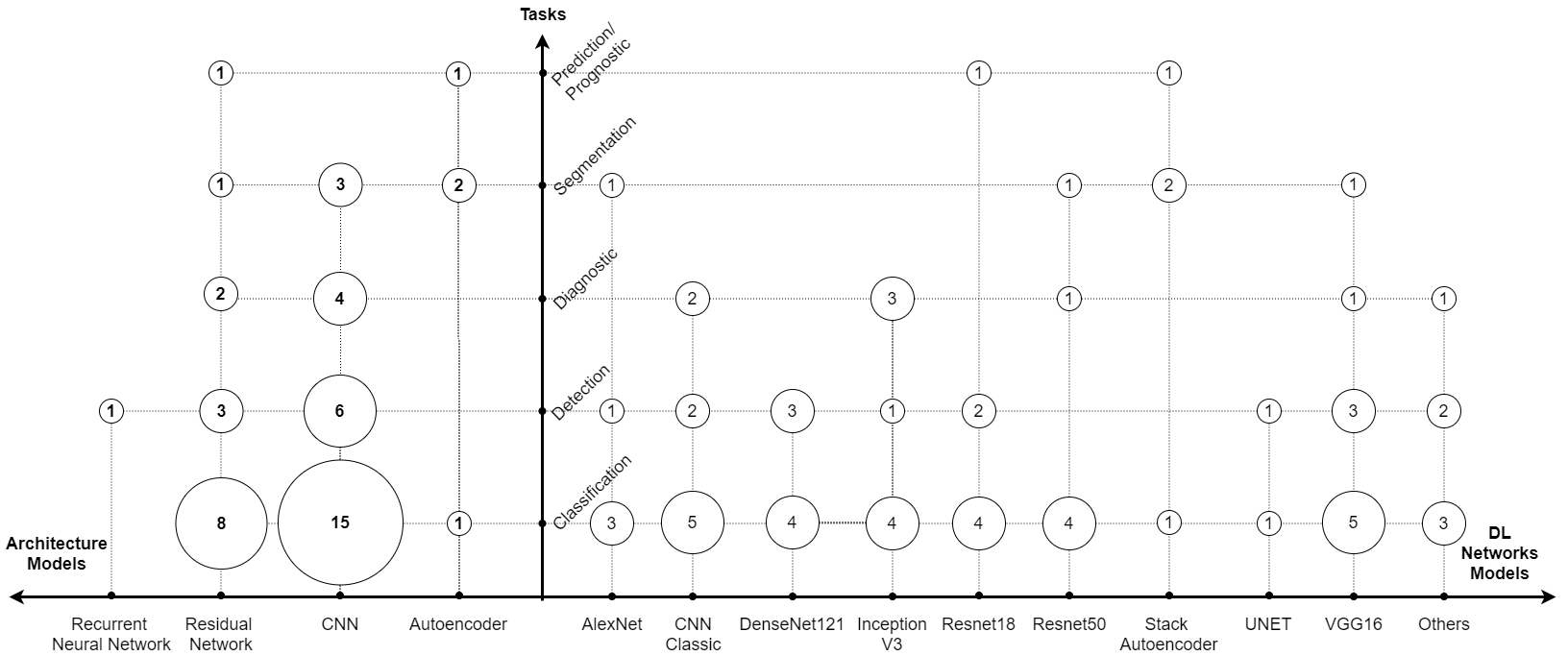}
    \caption{DL network architectures and models used in tasks. The value inside the bubble indicates the number of studies.}
    \label{fig:bubble_architetureVsModelsNN}
\end{figure*}

Both publications in journals and events are good sources of citation and provide many guidelines for future research. While journals are generally the most chosen periodicals, conferences are the most cited events, and in this research agenda, studies have strongly followed this path.


Journals are usually related to original research that has gone through a rigorous process with many rounds of peer expert review in the field. Magazines can bring opinions from authors and may not necessarily be supported by scientific literature, although they should not be overlooked in systematic studies. Event articles such conference and symposiums on the other hand have a faster and less rigorous review process and favor interaction with international audiences working in the same field, with negotiations and feedbacks being common.

\begin{figure}[hb]
    \centering
    \begin{tikzpicture}
    [   pie chart,
        slice type={A}{blue!70},
        slice type={B}{gray},
        slice type={C}{green!70},
        slice type={D}{orange!70},
        slice type={E}{yellow!70},
        pie values/.style={font={\small}},
        scale=2.18     ]
       \pie{}{48.6/A, 5.4/E, 13.5/C, 13.5/B, 18.9/D}
        \legend[shift={(1.25cm,1cm)}]{{Classification}/A, {Detection}/D, {Diagnostic}/B, {Segmentation}/C, {Prediction/Prognostic}/E}
    \end{tikzpicture}
    \caption{Division of tasks covered in the studies. It is important to note that some studies may cover more than one task.}
    \label{fig:tasks-studies}
\end{figure}
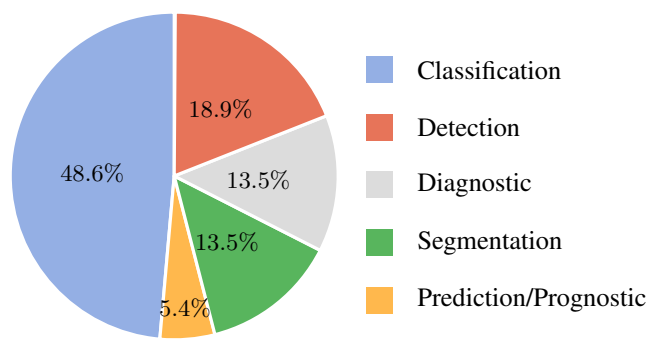

\vspace{.2cm}
\textbf{RQ2:} \textit{Which tasks applied to PCXR images are most addressed by DL techniques?}
DL techniques have been covering several tasks related to PCXR images as we have already seen in Table \ref{tab:DLarchitecturesVsTasks} and we can see in Figure \ref{fig:tasks-studies} some of these tasks have received more attention in cases of classification and detection, while prediction/prognostic tasks less attention. However, other tasks such as registration, recovery, imaging and improvement have been neglected by DL studies with PCXR and may be related to issues of data unavailability, human resources (for preparation, evaluation, validation) or the complexity of the task itself, to quote some, which opens space for new research.


\vspace{.3cm}
\textbf{RQ3:} \textit{Which neural network architectures are most used?}
The DL has a large number of architectural models that can be classified in several ways, such as the number of layers, training method or algorithm, or the type of network. However, the choice and adoption of one neural network architecture must consider another more important aspect, its application domain. Long Term Memory (LSTM), for example, are commonly applied in the understanding and translation of natural language, gesture recognition, and writing; Autoencoders to reduce dimensionality, adverse networks in resource learning and topic modeling, Residual Networks (ResNet) and Convolutional Neural Networks (CNN) for image recognition.

Thus, considering the scope of this research, the result presented in Table \ref{tab:DLarchitecturesVsTasks} was already expected, since the CNNs and ResNets architectures are specialized in image recognition and classification. They present excellent records of precision and accuracy and many of them are available in a open access format in several programming languages.  

In relation to the DL network models, the use of VGG16 used by 10 studies in 4 different tasks stands out as we can seen in Figure \ref{fig:bubble_architetureVsModelsNN} and the classic CNN and Inception V3 networks are also widely used. In relation to "Others" in this Figure, they refer to: Deep Neural Networks of Discriminant Analysis (DisAnDNNs), Deep Marginal Learning (MaShDL), R-CNN Mask, MobileNet-v2, Recurrent Scale Neural Network (SRCNN) and Siamese Convolutional Network (SCN). These DL architectures were mentioned only once each.


Finally, open questions from these studies point to other architectures that could be used, such as Adverse Generating Networks (GANs) to overcome data limitation by generating realistic false data, LSTM networks for generating reports in order to provide greater clarity and intelligibility to users, and RBM or Autoencoder networks for unsupervised training. Autoencoders were used in 4 studies (S1, S8, S20, and S21) but for other purposes. 



\begin{figure*}
    \centering
    \includegraphics[width=.99\textwidth]{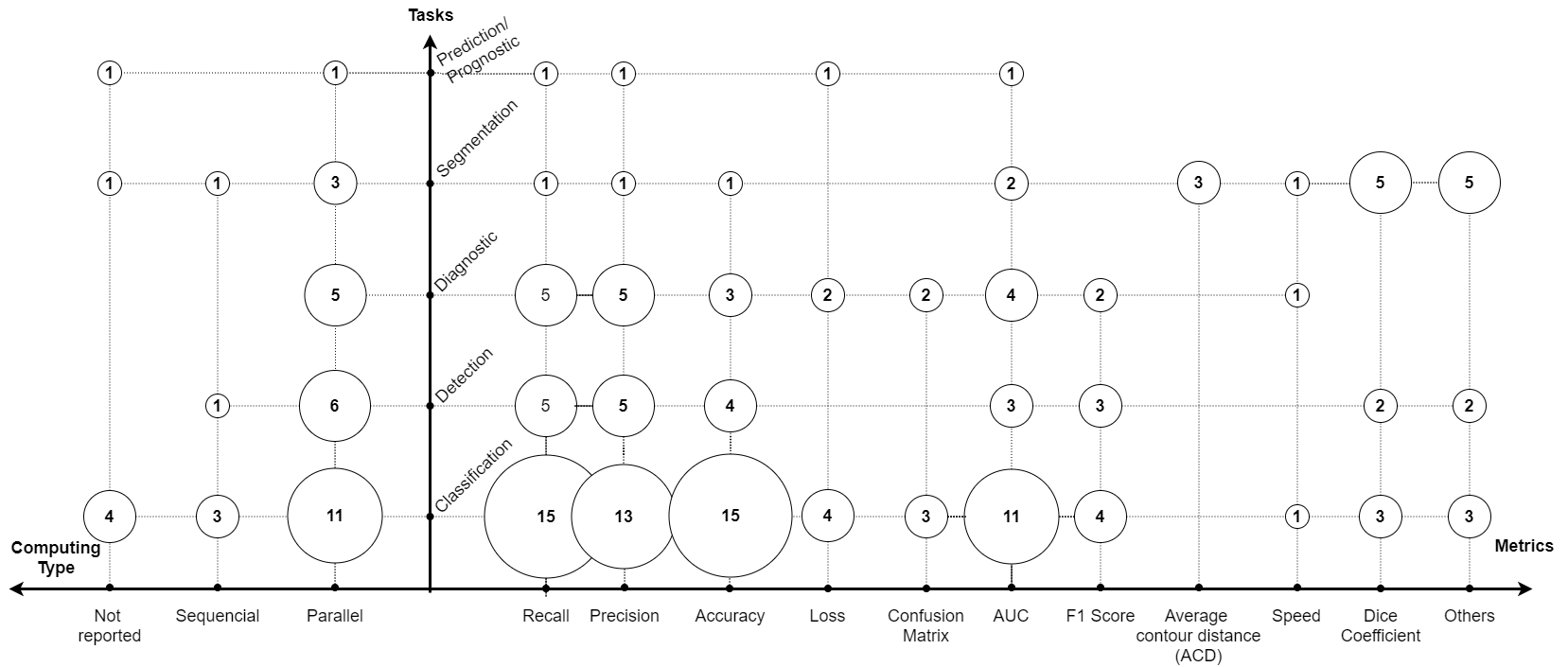}
    \caption{Processing type and metrics used in tasks. The value inside the bubble indicates the number of studies.}
    \label{fig:bubble_ProcessingVsMetrics}
\end{figure*}



\vspace{.3cm}
\textbf{RQ4:} \textit{Which metrics are used for assessment purposes?}

To answer this question, we can go back to Table \ref{tab:metricsUsed} and look at the variety of popular metrics used in the studies, or take a closer look at Figure \ref{fig:bubble_ProcessingVsMetrics}. These metrics are used together because the use of a single metric may not provide the complete measure of the task been solved. Dice coefficient was used by 100\% of studies that covered segmentation, while the recall, precision and accuracy metrics were almost absolute in the classification task. It shows that for application of DL in PCXR images these popular metrics have been satisfactory and are still the main choices.

It is also important to say that the choice of metrics is directly linked to the problem under analysis and that it is crucial that it is appropriate to guarantee the results and allow generalizations and comparisons, and that these metrics when associated with some method of cross-validation can guarantee greater robustness to the results, since it is common to find in studies solutions when applied to new data present very different results. 

Still in this regard, issues such as those referring, for example, to adversary attacks (which deceive the process of training and classifying a network by including noise), have also raised doubts about how much these popular metrics actually guarantee the generalization and robustness of solutions.

\vspace{.3cm}
\textbf{RQ5} \textit{How is each  DL technique in detail?} 
As mentioned in \textit{RQ4} and seen in the Figure \ref{tab:DLarchitecturesVsTasks}, CNNs and ResNets were the predominant architectures in the selected studies, being applied to the classification and detection tasks (mainly pneumonia). Regarding the models of these architectures, we can see from Figure \ref{fig:bubble_architetureVsModelsNN} that several models have been used, from classic CNN to AlexNet and UNET models. Therefore, considering only these main architectures, all studies used supervised training as a learning technique with the application of backpropagation and gradient descent based algorithms which are typical features of these architectures \cite{shresthaReviewDL2019}. 

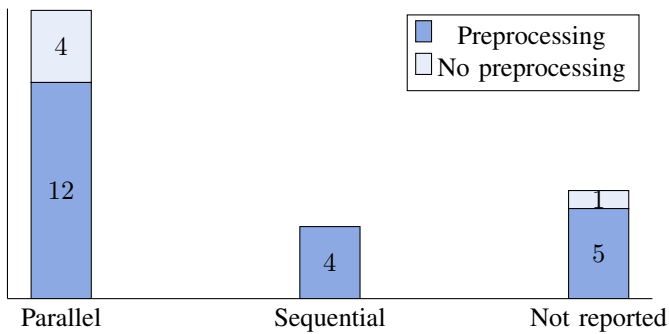
\begin{figure}[hb]
\begin{tikzpicture}
  \begin{axis}[
    ybar stacked, ymin=0, ymax=16.1, bar width=8mm, width= 0.56\textwidth, height = 0.3\textwidth,
    axis x line* = bottom, axis y line* = left, 
    ymajorgrids = true, ytick =\empty,
    symbolic x coords={Parallel,Sequential,Not reported},
    xtick=data, nodes near coords, 
  ]
  \addplot[style={fill=blue!75, mark=none}] coordinates { ({Parallel},12) ({Sequential},4) ({Not reported},5)};
  \addplot[style={fill=blue!15, mark=none}]  coordinates { ({Parallel},4) ({Sequential},0) ({Not reported},1)};
  \legend{Preprocessing, No preprocessing}
  \end{axis}
  \end{tikzpicture}
  \caption{Projection between preprocessing and computing type, the axis ’y’ indicating number of studies}
  \label{fig:preprocessingXapproachprocess}
\end{figure}

\begin{figure*}
    \centering
    \includegraphics[width=.99\textwidth]{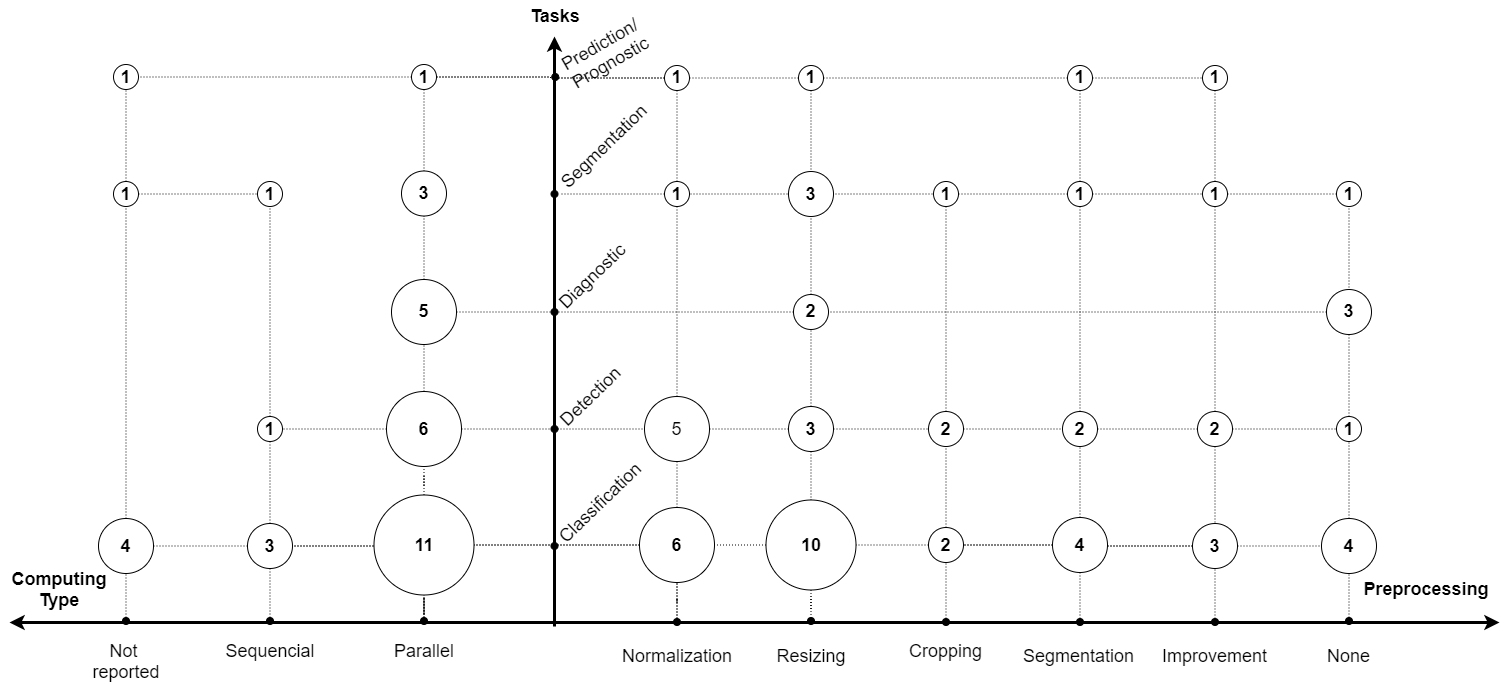}
    \caption{Processing type and preprocessing method used in tasks. The value inside the bubble indicates the number of studies.}
    \label{fig:bubble_ProcessingVsPreproc}
\end{figure*}

Regarding the computing type, the Figure \ref{fig:preprocessingXapproachprocess} derived from Table \ref{tab:PreprocessingVsComputationType} shows that parallel computing has become a standard approach, mainly due to advances in graphics processing units (GPU). Only four of the selected studies used sequential computing. Also, an important note is that most studies (80\% - 21/26) used preprocessing steps, especially those that used parallel computing (75\% - 12/16). Preprocessing steps as resizing and normalization are the most applied (Figure \ref{fig:bubble_ProcessingVsPreproc}) because of the benefits they can bring, since training networks is an expensive task and steps that reduce this work are usually mandatory.


\begin{figure}[ht]
    \begin{tikzpicture}
        [   pie chart, scale=2.2,
        slice type={legno}{red!50},
        slice type={caffe}{green!50},
        pie values/.style={font={\small}}   ]
        \pie{}{73/caffe,27/legno}
        \legend[shift={(-2cm,0.7cm)}]{{Private}/legno, {Public}/caffe}
    \end{tikzpicture}
    \caption{The ration of public and private PCXR datasets.}
    \label{fig:public_private-pcxr}
\end{figure}
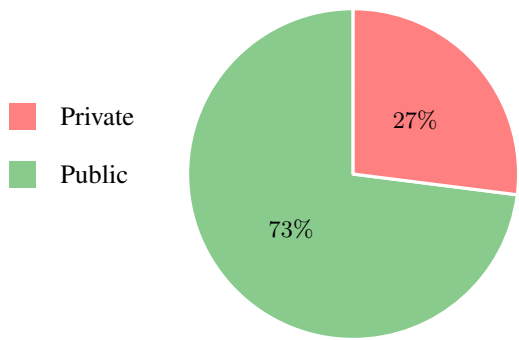

\vspace{.3cm}
\textbf{RQ6:} \textit{What  are  the  datasets  used,  and  how  are  they organized?}
Figure \ref{fig:graf-amount-datasets} and Figure \ref{fig:public_private-pcxr} show that only two sets of public data were cited, accounting to 73\% (19/26) of the studies. The first is the Guangzhou \cite{Kermany2018-Guangzhou} data set that was prepared specifically to hold PCXR images, while the second is a subset of NIH \cite{wang2017-NIH-dataset} data set which also contains adults X-ray images. This reveals that public data sets are important to improve this research field but it also shows that there is a scarcity of public data sets of PCXR images. Regarding the number of images, although these two data sets have a larger number of samples than the private data sets, this number is still considered small for neural network training standards, and therefore techniques such as data augmentation, pretraining or transfer learning, and fine-tuning are often used in most of these studies to compensated this reduced number of data. 

In this sense, although CXR is one of the most performed diagnostic tests in the world, this is not reflected in the number of available pediatric imaging data. Reasons for that include restrictions on the privacy and security of patient records and the difficulty to describe and label medical images. For example, the segmentation process is always the one that suffers most from the lack of data, because creating this type of data sets requires a lot of time, experience, and care to correctly define the image masks that will be used as a reference.


In addition, we point out that the number of images does not determine the quality of the study or the DL technique applied. Likewise, the increase in the number of images also has limitations, as noted in \cite{sun2017revisiting}. However, we cannot rule out that the greater the availability of data, the greater the probability of generalizing the solutions, and the less change in overfitting.

\vspace{.3cm}

\textbf{RQ7:} \textit{What  is  the  primary  type  of  contribution?} 

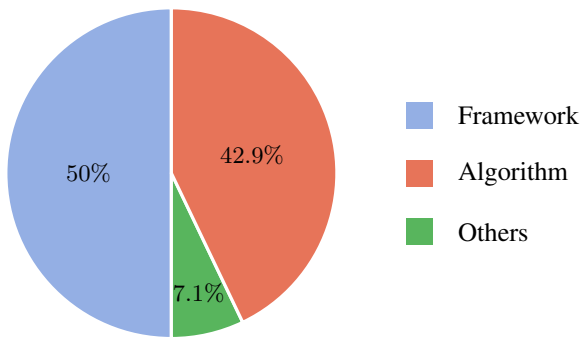
\begin{figure}[hb]
    \centering
    \begin{tikzpicture}
    [
        pie chart,
        slice type={comet}{blue!70},
        slice type={legno}{orange!70},
        slice type={caffe}{green!70},
        pie values/.style={font={\small}},
        scale=2.2
    ]
        \pie{}{50/comet,7.1/caffe,42.9/legno}
        \legend[shift={(1.5cm,0.7cm)}]{{Framework}/comet, {Algorithm}/legno, {Others}/caffe}
    \end{tikzpicture}
    \caption{Result proposed in the analyzed studies}
    \label{fig:resultByStudy}
\end{figure}

Most of the analyzed studies presented some framework or algorithm (see Figure \ref{fig:resultByStudy}) linked to an application or an adaptation of the CNN and ResNet architectures as a final result. Some of them combined these two architecture in a single one. The exceptions were the studies \textbf{S5} and \textbf{S13} in which they proposed study cases. The first evaluated the use of an automated DL software by health professionals with no coding experience or DL expertise. The second evaluated the hypothesis of CNN in-depth model automates the Brasfield score in CXRs of patients with cystic fibrosis (CF) and showed that the performance of their CNN model is similar to that of a pediatric radiologist.

Based on this topic of investigation, we can point out some conclusions: i) there is a great interest by researchers in using, improving and developing models of deep learning, ii) the works are unanimous in pointing out the data limitation as well as in emphasizing the preparation of the models to ensure the generalization of the results to external samples, i.e., the data used in the validation is different from those used in the construction of the models, and iii) although the researchers show excellent performance in the use of their models, none of them brought a product or application as a result, which can be interpreted as studies which are still in the experimental stage.

\vspace{.3cm}
\textbf{RQ8:} \textit{How can the research be classified?}
Based on Petersen et al. \cite{petersen2015guidelines} and Wieringa et al. \cite{Wieringa2006} and as noted in the Table \ref{tab:ResearchType&Method},  the classification that best defines the type of research of the analyzed studies (88,5\% of them) is the ``Evaluation Research'' while the method is mostly ``Controlled Experiments'', exception of \textbf{S14}, which is a ``Case Study''. This classification occurs because these studies propose new techniques or enhancements to existing techniques for solving tasks related to PCXR images. The authors of these studies also discuss their proposals and compare them with other related studies. Besides that, they are performed in an academic environment under specific conditions which means that they used a well-defined data set that is not affected by external factors.

Thus, after analyzing the \textit{RQ1-8} derived from our primary research question (PRQ) we can, based on these RQs, propose an answer to our PRQ.

\vspace{.3cm}
\textbf{PRQ:} \textit{What is state of the art on DL and PCXR-related tasks?}
\vspace{.1cm}


From the extraction carried out in the 26 studies, it is clear that there is significant interest from the scientific community in this area of research. Besides, this interest is increasing every year, as can be seen in the graph in Figure \ref {fig:grafYearVsVehicle}.


The publications are equally divided between journals and events and most are classified as evaluation research because they apply controlled experiments as a research method. This evidence added to the results presented in these studies allows us to see a constant process of evolution so that new limits are being established each day, leading to efficient, effective, and safe methods to be applied in hospital and health care environments.


The techniques applied in the selected studies allow us to know the main DL architectures applied to PCXR images. We observed that the Convolutional Neural Network (CNN) and the Residual Neural Network (ResNet) appear as the main actors in image processing applications, achieving high precision and accuracy in the analysis of the results.

In relation the tasks about PCXR images, some seem to attract more attention from the scientific community, such as classification and detection, but others have also been researched, which is a good indication of their importance in this research field. In addition, the small number of public PCXR data sets has been an additional challenge, currently circumvented with learning transfer and data augmentation techniques, but not yet explored using Adverse Networks, since the 26 selected articles did not investigate these architectures.


Although DL has shown impressive advances in many fields, in the medical field and more specifically in the analysis of CXR images, this technique can definitely be further explored. In the current context, we can say that applications are limited and, therefore, in the infancy stage. Many limitations still need to be overcome, such as the better readability of the models that allow the confrontation with the opinion of specialist doctors, the establishment of international standards and specific metrics to guide and validate the results of the studies and the transition of these proposed solutions, which are still in the research field, for application in industrial, commercial and hospital environments.


In summary, and already noticed by \cite{van2017fifty} and \cite{litjens2017survey}, DL is an excellent, powerful and ever-expanding expanding technique that can combine image analysis and radiology text reports analysis, for example. Thus, DL brings incredible possibilities that can make us believe that CAD systems that generate automated reports for CXR images will soon become reality.




\section{Research agenda}
\label{sec:researchAgenda}

In this research agenda, we carefully analyzed 26 studies [S1-S26] and found some evidence of gaps still present in the application of DL in pediatric X-ray images and in the state of the art of DL techniques. The existing solutions found in this research, although showing promising results, are still in the maturation stage with responses that are still fragile for real clinical applications. 
As a contribution, we have outlined a preliminary, but well-founded, research agenda to close this gap, which contains studies that:

\begin{enumerate}
    \item establish objective metrics for each task that helps researchers measure the performance and generalization of their solutions, as proposed by the American College of Radiology or even, that allow calculating the uncertainty estimates of the networks and their confidence level by physicans \cite{kendall2017uncertainties}, by \textbf{RQ4} we saw that popular metrics can be fragile in ensuring generalization of results;
    
    \item establish a set of standards for the creation and sharing of databases that are, for example, similar to the concept of ATM network or to the gold standard diagnosis and that guarantee security and anonymity, none of the analyzed studies explicitly mentioned this gap;
    
    \item considering the points raised by \textbf{RQ5}, evaluate the possible impact of reduced dimensionality (resizing) on the accuracy of the models. Although the absolute majority of the 26 studies analyzed perform this, no observation other than the processing cost is mentioned in this regard.
   
    
    \item considering the points raised by \textbf{RQ6}, assess the impacts of generating data annotations via crowd-sourcing, especially for those whose process requires a high level of specialization, brings fatigue and slowness, or is scarce and very expensive;
    
    \item investigate architectures DL based on more than just data, for example, models based on a combination of data and physics \cite{karpatne2017physics}, which can help with both generalization and interpretability issues, by \textbf{RQ2} we saw only the image processing being addressed;
    
    \item Demonstrate the robustness/fragility of DL architectures applied to PCXR against adversary attacks or the presence of external noise. None of the selected studies addressed this issue;
    
    
    \item investigate the application of DL in the task of generation, registering or retrieval pediatric chest X-ray images. Those tasks were not even mentioned in the analyzed studies as discussed in \textbf{RQ2};
    
    \item evaluate unsupervised DL models, such as variational auto-encoders (VAEs) and generative adversary networks (GANs), mainly to deal with unbalanced sets between classes, scarce in number or with unlabeled data, as we discuss in \textbf{RQ3};
    
    
    \item demonstrate a possible gain or not in the use of specific training as opposed to transferring learning in order to deal with the limited number of data sets with annotated images.
    
\end{enumerate}
Finally, this agenda goes beyond the simple quantitative investigation of DL techniques applied to PCXR images. It focuses on questions whose answers may have important implications for adopting or not adopting deep learning standards, strategies and architectures, as well as for lifting their limitations and pointing out their opportunities.

\section{Discussion}\label{sec:discussion}

Although there are numerous metrics for assessment and measurement DL models, there is no international reference manual or standard for measuring and evaluating deep learning tasks, especially those associated with pediatric chest X-ray images, as far as we know. A trend may be the adoption of assessment standards such as those applied to major public challenges as in Kaggle competitions or initiatives like that of the American College of Radiology (ACR) through algorithm review processes \cite{moore2019machine}.


Due to the lack of standard metrics, performance comparisons between related work are difficult to be made, consequently, DL in PCXR is incipient to support medical decisions. In our study, none of the analyzed studies presented an application for clinical medicine, even in those with expressive results. A single case study \textbf{(S14)} was reported which indicates that the solutions are not sufficiently mature or fully reliable for real-world applications.

In contrast, DL has been showing impressive results in several research fields, surpassing, in some cases, human performance, as in precision agriculture, in autonomous vehicles and in game industry. In the field of medical imaging, DL has been important in the analysis of optical coherence tomography (OCT) for the detection of diabetic retinopathy, for example. On the other hand, DL applied to PCXR images, seems incipient and with less impact. In our view, applications of this type need further testing in order to ensure greater generalization and robustness of their solutions.


As we presented, our SLM is not a list of all the open questions about the application of DL in pediatric chest X-ray images, and what we pointed out are only guidelines that can be followed in future works (Section \ref{sec:researchAgenda}). Therefore, our discussion tries to bring out the most latent aspects presented in the 26 selected studies in order to point out directions to guide other researchers in their studies.


In this sense, our question about the maturity of the application of DL in pediatrics CXR images is provocative and, therefore, we do not intend here to give a definitive answer. We only provoke some questions that can help the reader to make their own experiments and investigations.


We believed that the application of DL in pediatric chest X-ray images is still in childhood because although the DL was already applied to the analysis of medical images in 1995 by Lo et al. \cite{lo1995artificial}, only from 2015 it has a massive application in medical images \cite{van2017fifty} and more precisely in 2016 in PCXR images \textbf{(S8)}. Therefore, if we consider only this last information, we could chronologically define DL in PCXR images as a 4-year-old child. Similar observation to that of LeCun et al. \cite{lecun2015deep} which states that systems that combine DL and reinforced learning are still in their childhood although they outperform passive vision systems in classification tasks.

As already discussed before, another point that may explain the maturity stage of DL applications in CXR pediatrics is the fact that it is supported by supervised learning so that it depends on a large amount of data and participation of professionals in an expensive data labeling process. It also explains why the selected papers work with controlled experiments \textbf{(RQ8)} and why clinical products or applications were not presented. In this sense, the generalization of the results is fragile and requires effective metrics for the evaluation of the performance of the proposals. Furthermore, it requires interpretability, simplicity, and mathematical explanation about the decisions of the models in order to enable the debate with doctors and discuss ethical questions about responsibility in these matters.


Despite of that, we can not ignore the fact that the application of DL in pediatric CXR images has surpassed the state of the art in several tasks and has allowed obtaining impressive results, equivalent or superior in some cases to those achieved by specialist doctors. Thus, results of this type reflect the maturation of the application of DL in PCXR images, mainly in the development of tools to support medical professionals, such as those related to the tracking of suspected cases, suppression of bone structures, pre-processing processes for archiving and improvement storage, orientation correction or CXR vision classification.

At last, CNN and ResNet are the two DL architectures more currently used in PCXR images investigation. Also, tasks such as classification, detection, and segmentation have received more attention than other issues as acquisition and image registration. We note that this field of research is on the rise and has received attention from the scientific community so that it leads us to believe that it is only a matter of time for other DL solutions to be applied to PCXR images as well. Our bet is that in a short time we will see annotation and report generation, strong use of unsupervised learning, crowdsourcing to deal with data scarcity of labeled data and unbalanced classes, and parallel computing to work with high image resolutions and dimensions.

\section{Conclusions}
\label{sec:conclusion}


Deep learning is not a new topic since it was applied more than three decades ago for the first time. The analysis of chest radiographs using computer techniques have also been investigated for a long time \cite {van2017fifty}. Deep learning and X-ray images motivated many works, and the combination of both has leveraged many academic papers (Figure \ref{fig:process-sms}), including several secondary studies \cite{lee2019deep, feng2019deep, yasaka2018deep, van2017fifty, pande2016computer}. In this way, these published work bring snapshots of different moments in the evolution of these themes. However, no SLM study has been conducted to investigate research related to the development of deep learning techniques applied to pediatric images. Therefore, we presented a systematic mapping that brings a deep immersion of the details and characteristics of deep learning techniques applied to PCXR images. Also, we presented a research agenda to meet the gaps and trends of this topic in order to add value to other research.




In our work, from 178 studies, we selected the 26 most relevant. Then, we applied in these 26 primary studies a complete analysis for extracting different data for discussion. We noticed that it is a hot topic in which the interest in it is growing progressively. However, the absence of a similar study, as far as we know, justified the design of this unprecedented research agenda. In this regard, the comprehensive protocol presented in Section \ref{sec:sdudy-protocol} allows other researchers to validate, reproduce, and extend this study. This organization and detailed descriptions are some of our main contributions.



We conclude that the investigated area is on the rise because the 26 selected works showed important results of the application of computing in the medical field. In addition, these studies open the way for other studies in which image analysis can contribute to the diagnosis and clinical evaluation and can even contribute to saving lives. In this sense, although there is a lot of work to be done in order to leverage the applications of deep learning in pediatric chest X-ray analysis, we are sure that this research agenda has contributed to the growth and maturation of this research area.

\section*{Acknowledgments}

We thank all those who collaborate directly and indirectly for the execution of this study, in particular to the deep learning specialists of the group Deep Learning Brazil who collaborated in the definition of several keywords of our search string.

\vspace{.75cm}
 \bibliographystyle{IEEEtran}
 \bibliography{references}

\end{document}